\documentclass[preprint2]{aastex}
\usepackage{graphicx}
\usepackage{epsfig}
\usepackage{lscape}
\usepackage{color}

\begin{document}
\slugcomment{}

\shorttitle{}

\title{Search for pulsed $\gamma$-ray emission from globular cluster M28}
\shortauthors{Wu et al.}

\author{J. H. K. Wu\altaffilmark{1}, C. Y. Hui\altaffilmark{2}, E. M. H. Wu\altaffilmark{3},  A. K. H. Kong\altaffilmark{1,4}, R. H. H. Huang\altaffilmark{1}, P. H. T. Tam\altaffilmark{1} , J. Takata\altaffilmark{3}, K. S. Cheng\altaffilmark{3}}

\altaffiltext{1}
{Institute of Astronomy and Department of Physics, 
National Tsing Hua University, Hsinchu, Taiwan }
\altaffiltext{2}
{Department of Astronomy and Space Science, Chungnam National University, Daejeon, Republic of Korea}
\altaffiltext{4}
{Golden Jade Fellow of Kenda Foundation, Taiwan}
\altaffiltext{3}
{Department of Physics, University of Hong Kong, Pokfulam Road, 
Hong Kong}

\email{cyhui@cnu.ac.kr \& wuhkjason@gmail.com}


\begin{abstract}
Using the data from the Large Area Telescope
on board the \emph{Fermi} Gamma-ray Space Telescope, 
we have searched for the $\gamma$-ray pulsations from the direction of globular cluster
M28 (NGC 6626). We report the discovery of a signal with the frequency consistent with 
that of the energetic millisecond pulsar (MSP) PSR~B1821-24 in M28. A weighted H-test 
test statisic (TS) of 28.8 is attained which  corresponds to a chance probability of 
$\sim10^{-5}$ ($4.3\sigma$ detection).
With a phase-resolved 
analysis, the pulsed component is found to contribute $\sim25\%$ of the total observed 
$\gamma$-ray emission from the cluster. On the other hand, the unpulsed level provides a constraint for the 
underlying MSP population and the fundamental plane relations for the scenario of inverse Compton 
scattering. Follow-up timing observations in radio/X-ray are encouraged for further investigating 
this periodic signal candidate. 
\end{abstract}

\section{INTRODUCTION}
The first millisecond pulsar (MSP), which is a rejuvenated old neutron star through accreting
matter from its companion, was discovered 30 years ago (Backer et al. 1982).  
It has long been suggested that they are the descendants of low-mass X-ray binaries 
(LMXBs; Alpar et al. 1982). In comparison with the Galactic field, the formation rate per unit mass
of LMXBs in globular clusters (GCs) is orders of magnitude higher because of the frequent stellar 
encounters (Katz 1975; Clark 1975; Pooley et al. 2003; Hui, Cheng \& Taam 2010). Therefore, 
it is not surprising that GCs should host a large population of MSPs. 
Since the first cluster MSP, PSR~B1821-24, has been discovered in M28 (Lyne et al. 1987), 
dedicated radio pulsar surveys towards different clusters have resulted in the currently known 
population of 144 MSPs in 28 GCs.\footnote{see http://www.naic.edu/$\sim$pfreire/GCpsr.html for 
updated information.}
 
Since the launch of \emph{Fermi} Gamma-ray Space Telescope, a new population of 
$\gamma-$ray emitting globular clusters (GCs) have been detected (Abdo et al. 2010a; 
Tam et al. 2011). As MSPs are the only known steady $\gamma$-ray sources in GCs, they are 
suggested to be the contributors for the observed emission. The $\gamma$-rays from 
a GC are interpreted as the collective contribution from the entire pulsar 
population resides in it (Abdo et al. 2010a; Cheng et al. 2010; Hui et al. 2011). For most of the cluster MSPs, 
detecting the $\gamma$-ray pulsations is very challenging as the $\gamma$-ray flux
of individual pulsar is weak. To exacerbate the situation, the spatial resolution of 
\emph{Fermi} Large Area Telescope (LAT) does not allow individual MSPs in a GC to be resolved. This
results in a high background that make the pulsation search for any individual MSP very difficult. 
However, there are two notable exceptions, namely the GCs NGC~6624 and M28. 
Both of these GCs contain a very energetic and young MSP so that they can possibly stand out from 
the background.

For PSR~J1823-3021A in NGC~6624 ($P=5.44$~ms), its large spin-down rate, 
$\dot{P}=3.38\times10^{-18}$~s~s$^{-1}$, implies that it is the 
youngest MSP ($\tau\sim25$~Myrs) ever detected (Freire et al. 2011). 
Its spin-down luminosity is $\dot{E}=8.3\times10^{35}$~erg~s$^{-1}$, 
which is $\sim1-2$ orders of magnitude greater than the typical MSPs in GCs 
(Bogdanov et al. 2006). Thanks to the accurate timing model provided 
by the dedicated radio observations, its 
$\gamma-$ray pulsations has been revealed by ${\it Fermi}$ (Freire et al. 2011). 
Through a phase-resolved analysis, it has been shown that this single
pulsar dominates all the observed $\gamma-$rays from NGC~6624 (Freire et al. 2011). 
The $\gamma-$ray conversion efficiency of PSR~J1823-3021A, $L_{\gamma}/\dot{E}\sim0.1$, 
is found to be comparable with other $\gamma-$ray detected MSPs (Abdo et al. 2010b),
where $L_{\gamma}$ is the $\gamma-$ray luminosity.

PSR~B1821-24 in M28 (hereafter M28A) is very similar to PSR~J1823-3021A in many aspects. 
Its period ($P=3.05$~ms) and spin-down rate ($\dot{P}=1.61\times10^{-18}$~s~s$^{-1}$) 
imply its age and spin-down power to be 
$\tau\sim30$~Myrs and $\dot{E}=2.2\times10^{36}$~erg~s$^{-1}$ respectively, 
which makes it as the most energetic MSP has been found so far (Bogdanov et al. 2011). 
Together with its non-thermal X-ray spectrum, 
its sharp and narrow X-ray pulse profile strongly indicate that 
most of the observed X-rays from M28A are originated from the magnetosphere.
$\gamma-$ray emission from M28 has been detected by {\it Fermi} LAT (Abdo et al. 2010a). 
As its $L_{\gamma}$ is only a fraction of the spin-down power of M28A, 
it is possible that this pulsar can have a significant contribution to the observed $\gamma-$rays.
Together with its relatively short distance, 5.5~kpc (cf. Harris 1996; 2010 version), 
the cluster M28 is a promising target for searching $\gamma-$ray pulsation. 
In this Letter, we report our recent search for the possible pulsation from 
this GC by using {\it Fermi} LAT data.

\section{DATA ANALYSIS AND RESULTS}
In this work, we used the {\it Fermi} LAT data between 2008 August 04 and 2012 January 31. 
For the data analysis, the {\itshape Fermi} Science Tools v9r23p1 package, 
available from the {\itshape Fermi} Science Support Center\footnote{
http://fermi.gsfc.nasa.gov/ssc/data/analysis/software/}, was used. We used Pass 7 data and  
selected events in the ``Source" class (i.e.~event class 2) only. 
In addition, we excluded the events with zenith angles larger than 100$\degr$ to 
greatly reduce the contamination by Earth albedo gamma-rays. The instrumental response functions (IRFs) 
``P7SOURCE\_V6" were adopted throughout the study. Events were selected within a circular region-of-interest (ROI) 
with a diameter of $10\degr$ centered at the optical center of M28. 
Photon energies are restricted in the range of 200~MeV$-$300~GeV. 
This set of cuts is adopted throughout this work.

To investigate the spectral characteristic of M28 with the updated IRFs and background model, we performed 
an unbinnned likelihood analysis with the aid of \textit{gtlike} by assuming a point source with power-law 
with exponential cutoff (PLE) of the form $dN/dE\propto E^{-\Gamma}{\rm exp}\left(-E/E_{\rm cutoff}\right)$ 
at the nominal position of M28, where $\Gamma$ and $E_{\rm cutoff}$ are the photon index and the cutoff energy 
respectively. 
For modeling the background, we included the Galactic diffuse model ({\tt gal\_2yearp7v6\_v0.fits}),  
the isotropic background ({\tt iso\_p7v6source.txt}), as well as all point sources reported in 
the 2FGL catalog within $10\degr$ from the center of the ROI. All 
these 2FGL sources were assumed to be point sources which have specific spectrum 
suggested by the 2FGL catalog (Nolan et al. 2012). While the spectral parameters 
of the 2FGL sources locate within the ROI 
were set to be free, we kept the parameters for those lying outside our adopted ROI
fixed at the values given in 2FGL (Nolan et al. 2012). 
We allowed the normalizations of diffuse background components to be free. 
The best-fit PLE model is characterized by  
$\Gamma=0.96\pm0.22$ and $E_{\rm cutoff}$= 1.41$\pm$0.3 GeV with a test-statistic (TS) 
value of 825 which is highly significant. We have tested the robustness of the spectral results by 
repeating the analysis with different size of ROI. The fitted parameters from independent analysis are consistent 
within $1\sigma$ uncertainties.
In this model the photon flux between 200 MeV \& 300 GeV 
was found to be $(2.39\pm0.22)\times10^{-8}$~cm$^{-2}$~s$^{-1}$. 
The corresponding integrated energy flux is 
$f_{\gamma}=(3.17\pm0.29)\times10^{-11}$~erg~cm$^{-2}$~s$^{-1}$.
The spectral results are consistent with those reported by Abdo et al. (2010a) within $1\sigma$ uncertainties. 

\begin{figure*}[b]
\centerline{\psfig{figure=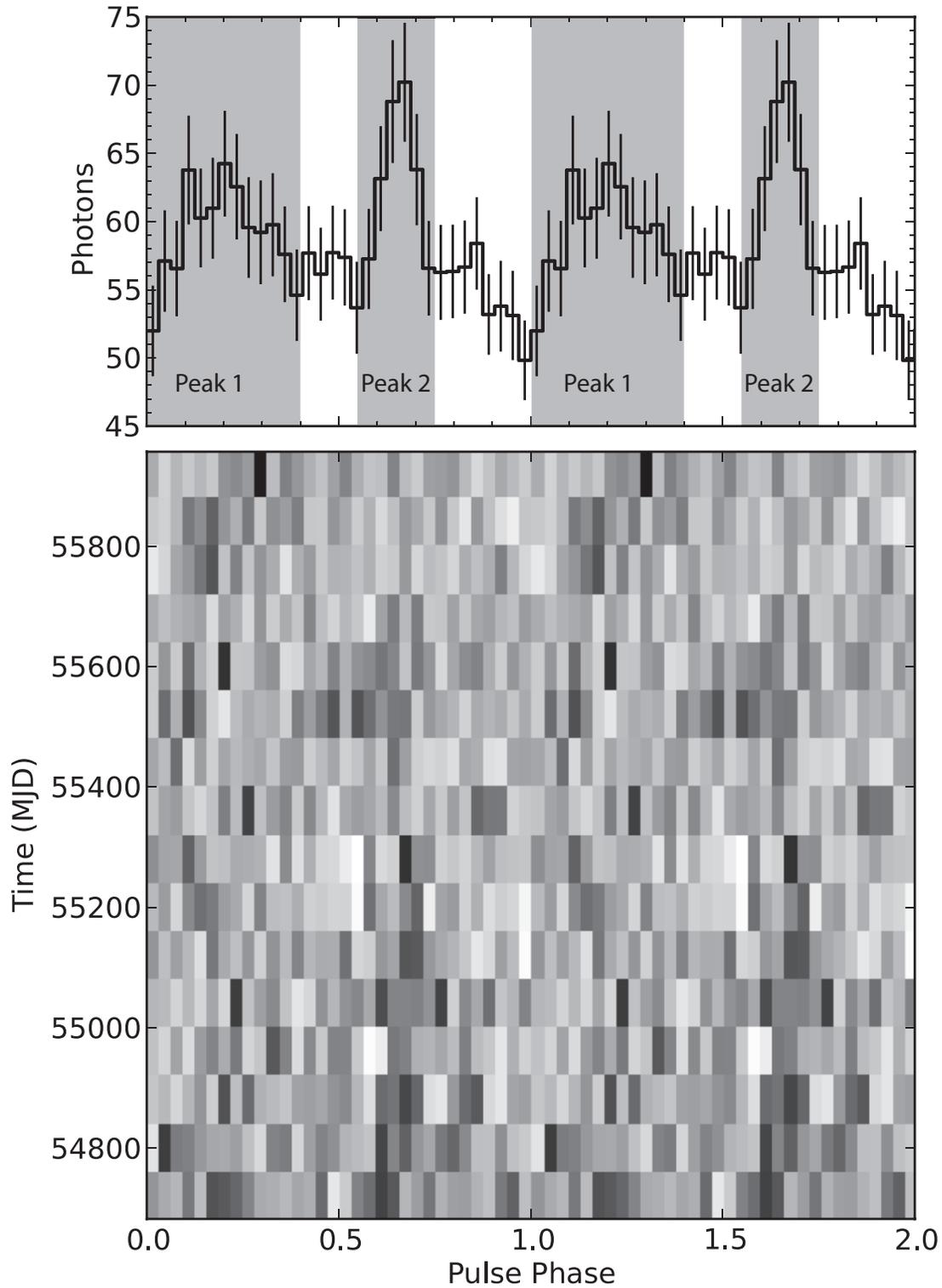, width=16cm,clip=,angle=0}}
\caption[]{\footnotesize
{\it Fermi} LAT $\gamma$-ray weighted light curve ({\it upper panel}) and the phaseogram ({\it lower panel}) of M28A.
A weight was assigned to each photon with the probability that it
comes from M28 by using the task {\itshape gtsrcprob}
in {\it Fermi} Science Tool. Two periods of rotation with a resolution of 40 phase bins per period is shown for clarity. 
The error bars of the light curve represent $1\sigma$ Poisson uncertainties. The shaded regions define the on-pulse intervals 
for Peak 1 and Peak 2.}
\label{ray_weighted_lc}
\end{figure*}

Assuming M28A is major contributor for the $\gamma$-rays from M28, 
we search for the possible $\gamma$-ray pulsation from this GC. We started by adopting 
the timing ephemeris determined for M28A from a long term observation with {\it Rossi} X-ray Timing Explorer
(Ray et al. 2008), which are tabulated in Table~1. 
 For barycentric correction, we used the updated planetary ephemeris JPL DE405
throughout this analysis. 
Following the method proposed by Kerr~(2011), we used the
best-fit model resulted from the phase-averaged likelihood analysis as described above for
assigning weight to each $\gamma$-ray photon by computing the probability that it originates from M28.
This enables us to alleviate the problem of source confusion with a more efficient background rejection.
We then assign a pulsar spin phase to every gamma-ray photons with energies $>$0.2 GeV and within 5$^{\circ}$ 
from M28A's direction (see Tab.~1). 
A promising signal with a weighted H-test TS of 28.8 has been found by directly folding up the data with this ephemeris
(de Jager \& B\"{u}sching 2010). 
The folded $\gamma-$ray pulse profile and the phaseogram (i.e. pulse phase as a function of time) with
the weighted photons are shown in the upper panel and the lower panel of Figure~\ref{ray_weighted_lc} respectively.

\begin{figure*}[b]
\begin{center}
{\psfig{figure=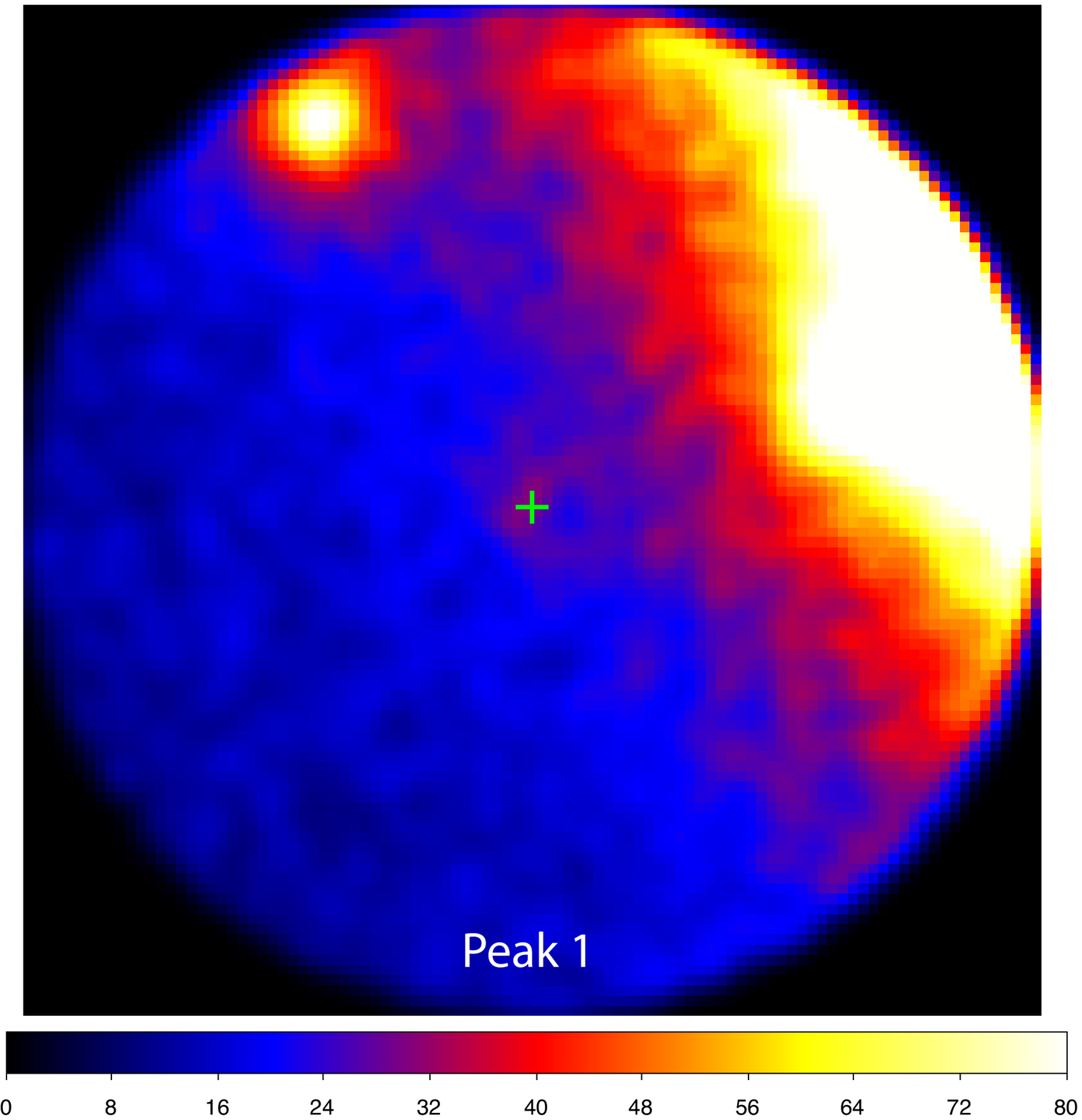, width=6cm,clip=,angle=0}}{\psfig{figure=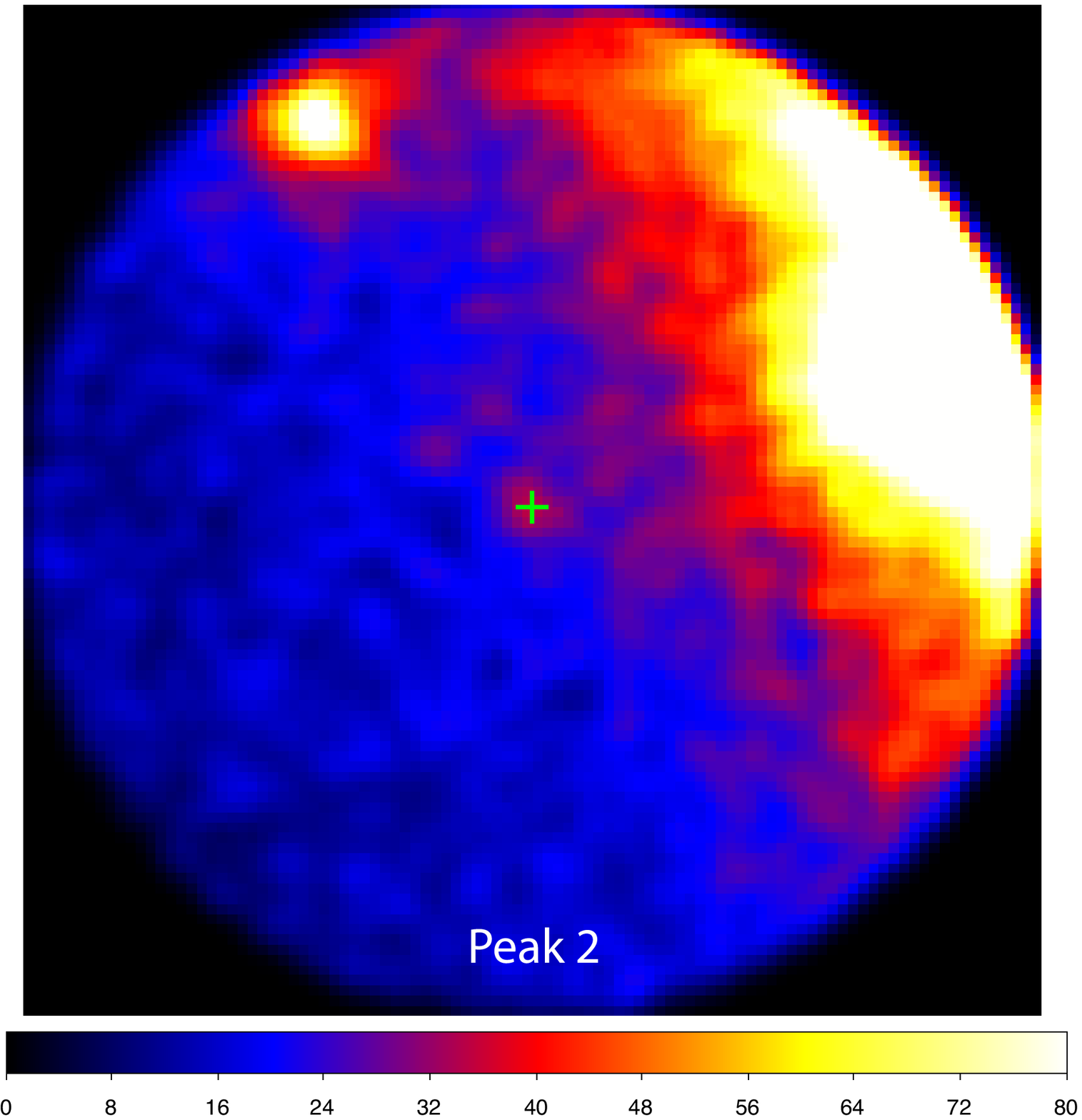, width=6cm,clip=,angle=0}}{\psfig{figure=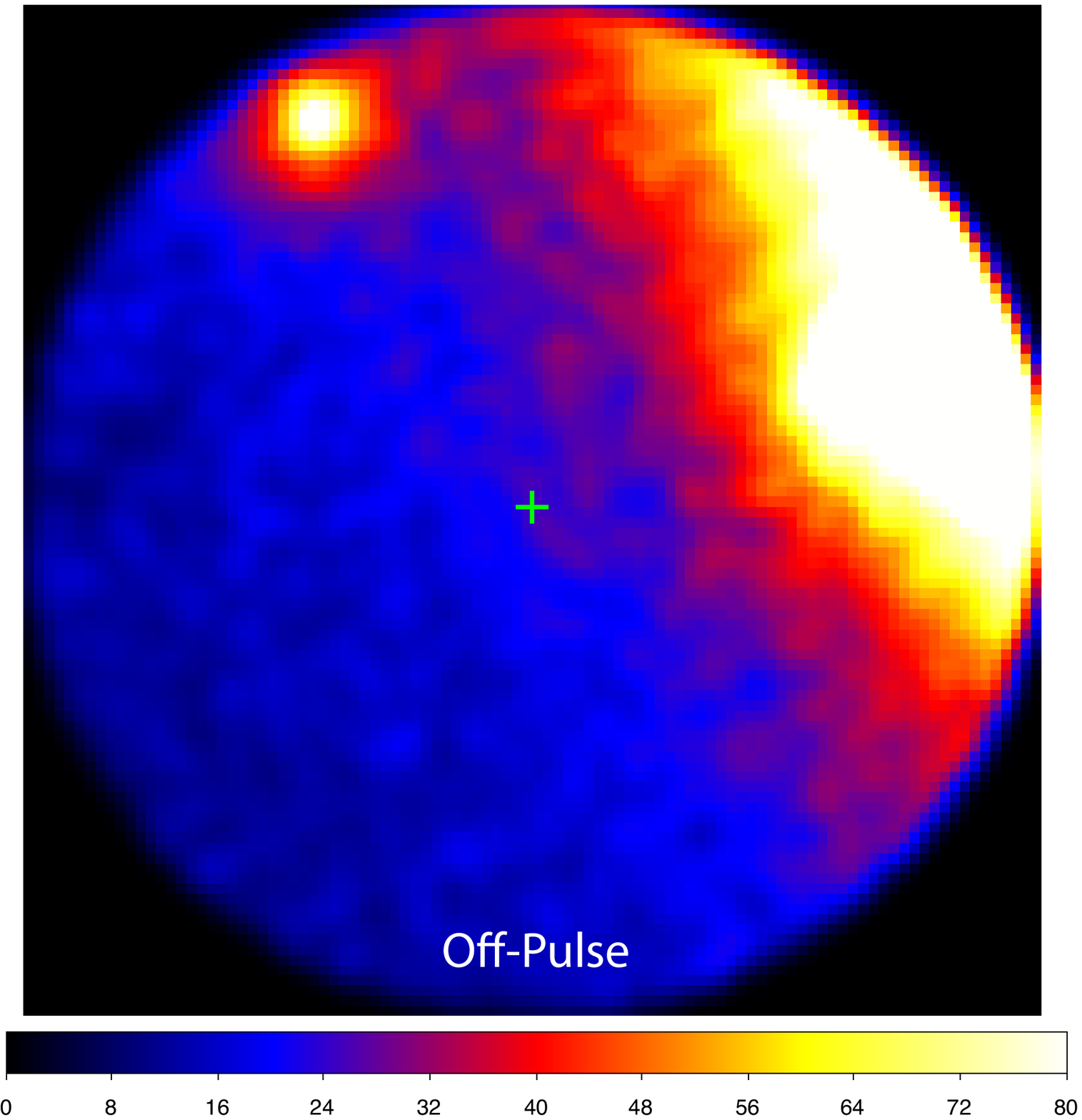, width=6cm,clip=,angle=0}}
\end{center}
\caption[]{\footnotesize {\it Fermi} LAT phase-resolved 
$\gamma$-ray count maps for events $>$0.2 GeV within 5$^{\circ}$ of the timing position of M28A 
(illustrated by the yellow cross). Top is north and left is east. 
The scale bar below shows the color scale of counts/pixel divided by the relevant phase interval.
{\it Left panel}: Peak~1 region (i.e. 0.0$<\phi<$0.4). 
{\it Middle panel}: Peak~2 region (i.e. 0.55$<\phi<$0.75). 
{\it Right panel}: Off-pulse region (i.e. 0.4$<\phi<$0.55 \& 0.75$<\phi<$1). 
The point source locates in the northeast of these maps 
is 2FGL~J1833.6-2104 and the bright extended emission in the northwest is due to the diffuse $\gamma-$ray emission from 
the Galactic plane.}
\label{countmap}
\end{figure*}

According to Figure~\ref{ray_weighted_lc}, it appears to have two peaks with one broader than the other. 
We then define the phase-interval for the peak~1 and peak~2 
to be $0-0.4$ and $0.55-0.75$ respectively. The rest is defined as the off-pulse component. With this 
definition, we show the {\it Fermi} LAT count maps of the sky region around
M28 at different phases in Figure~\ref{countmap}.
During the on-pulse intervals, a point-like $\gamma-$ray
source can be clearly seen at the pulsar position which is illustrated by the
yellow cross. On the other hand, a faint diffuse excess is 
found in the off-pulse phase (i.e. 0.40$<\phi<$0.55 and 0.75$<\phi<$1.0).
However, the limited photon statistic does not allow us to constrain the extent 
of this putative feature. 

This on-and-off nature of the $\gamma-$ray emission from M28 provides a strong support 
for the presence of periodic signal and leads us to a more detailed investigation. 
For investigating the possible spectral variations among peak~1, peak~2 and the unpulsed component, 
we performed a phase-resolved likelihood analysis. We have fitted their spectra with 
both simple power-law (PL) of the form $dN/dE\propto E^{-\Gamma}$ 
and PLE. The results are summarized in Table~2. According to the PL fits, there is no obvious 
change of the spectral steepness. We notice that the 
likelihood analysis that incorporate the PLE model results in a higher TS for all three 
components. For peak~2, we found that the spectral parameters for the PLE fit cannot be properly constrained. Therefore, 
we fixed the photon index at the value inferred in the phase-averaged analysis (i.e. $\Gamma=0.96$). 
Same as the cases of the PL fits, within the tolerence of the statistical 
uncertainties, we do not find any conclusive evidence for the spectral variation across 
the phase. Assuming the off-pulse component has a constant contribution across the whole
phase, $\sim75\%$ of the total observed flux is originated from this component.

\section{SUMMARY \& DISCUSSION}
In this Letter, we report our detection of $\gamma-$ray pulsation from the direction of the GC M28.  
We have found a periodic signal which is presumably originated from its energetic MSP M28A. 
Based on our phase-resolved analysis, the pulsed component contributes $\sim25\%$ of the total observed $\gamma-$rays. 
At a distance of $d=5.5$~kpc, this implies an on-pulse luminosity of 
$L_{\gamma}=4\pi d^{2}f_{\Omega}f_{\gamma}\sim3\times10^{34}f_{\Omega}$~erg~s$^{-1}$, 
where $f_{\Omega}$ is the fraction of the sky covered by the $\gamma-$ray beam. 
Assuming the pulsed emission is originated from M28A, this suggests a $\gamma-$ray conversion efficiency of
$L_{\gamma}/\dot{E}\sim0.01f_{\Omega}$. Some of the MSPs in the Galactic field, such as PSRs~J2124-3358 
and J0437-4715, have their $L_{\gamma}/\dot{E}$ found at this level (Abdo et al. 2010b). However, 
this is lower than $L_{\gamma}/\dot{E}\sim0.08$ as derived from the nearby MSPs (Abdo et al. 2009). 
If one adopt this as the intrinsic $\gamma-$ray conversion efficiency of M28A, this might suggest the observed 
period derivative is largely dominated by the acceleration of the pulsar along the line-of-sight due to the 
gravitational field of the cluster. 

On the other hand, the off-pulse luminosity is found at the level of $L_{\gamma}\sim8\times10^{34}$~erg~s$^{-1}$.
This estimate is useful for constraining the collective properties of the rest of the MSP population in M28. For 
explaining the unpulsed $\gamma-$ray emission from GCs, there are two main theories. One stream interprets the 
$\gamma-$ray emission from a GC is originated from the collection of the magnetospheric radiation from the entire 
MSP population resides in it (Abdo et al. 2010a; Venter et al. 2008,2009). Assuming an average spin-down power 
of $\left<\dot{E}\right>\sim2\times10^{34}$~erg~s$^{-1}$ and a characteristic conversion efficiency of 
$\sim0.08$, the off-pulse luminosity enables us to estimate the number of the rest MSP population
to be $N_{\rm MSP}\sim50$. This suggests that about one fifth of the underlying population has already been uncovered. 

Apart from the aforementioned standard scenario, inverse Compton scattering (ICS) between the relativistic pulsar 
wind particles and the ambient soft photons has also been proposed as another possible explanation for the origin of the 
$\gamma-$ray from GCs (Bednarek \& Sitarek 2007; Cheng et al. 2010; Hui et al. 2012). Cheng et al. (2010) 
found that the observed $\gamma$-ray spectra of GCs can generally be well-modeled by ICS between the $e^{-}/e^{+}$ 
in the pulsar wind of the whole MSP population in a GC and the Galactic background IR photons or starlight. 
And the two-dimensional 
regression analysis further suggests $L_{\gamma}$, energy density of the background optical/IR photon field and the 
stellar encounter rate/metallicity span a set of fundamental planes (Hui et al. 2012). The unpulsed level inferred for
M28 can be used to discriminate which relation(s) can better predict the collective contribution. Using the best-fit parameters 
for these fundamental plane relations (Equations~1-4 and Table~3 in Hui et al. 2011) and the updated 
GC parameters (Harris 1996; 2010 version), the relations involve metallicity 
and optical/IR energy densities result in an estimate in a range of $\sim\left(8-9\right)\times10^{34}$~erg~s$^{-1}$ which 
is consistent with the observed off-pulse luminosity. On the other hand, the best-fit relations that involve the encounter rate 
result in an estimate of $\sim2\times10^{35}$~erg~s$^{-1}$ which apparently overshoot the observed value. 

For further investigating this putative periodic signal, multi-wavelength observations are certainly required. In particular, the 
phase-aligned X-ray/$\gamma-$ray pulse profile will provide an important constraint for the high energy emission model. However, no existing
X-ray timing data is available for M28A in \emph{Fermi} era. As the timing noise of M28A is quite strong in comparison with 
other MSPs and it possibly exhibited glitches, the phase-alignment of multi-wavelength light curves subjects to a lot of uncertainties.
Therefore, follow-up timing observations in other wavelengths are encouraged for further investigations.  

\clearpage
\begin{table*}
\begin{center}
\caption[]{Ephemeris of PSR~J1824-2452A adopted from Ray et al. (2008).\\ }
\begin{tabular}{ll}
\hline\hline
\multicolumn{2}{l}{Parameter} \\ 
\hline
Pulsar name\dotfill & J1824-2452A \\ 
Right ascension, $\alpha$\dotfill & 18:24:32.00790550 \\ 
Declination, $\delta$\dotfill & -24:52:10.8076448 \\ 
Pulse frequency, $\nu$ (s$^{-1}$)\dotfill & 327.4056060517495439 \\ 
First derivative of pulse frequency, $\dot{\nu}$ (s$^{-2}$)\dotfill & $-$1.735361869603$\times 10^{-13}$ \\ 
Epoch of frequency determination (MJD)\dotfill & 53800 \\ 
Epoch of position determination (MJD)\dotfill & 53800 \\ 
Solar system ephemeris model\dotfill & DE405 \\
Time system \dotfill & TDB \\
\hline
\end{tabular}
\end{center}
\label{table1}
\end{table*}

\clearpage

\clearpage
\begin{center}
\begin{deluxetable}{lccc}
\tablewidth{0pc}
\tablecaption{Phase-resolved spectroscopy of M28A}
\startdata
\hline\hline
          & Peak~1 & Peak~2 & off-pulse component \\\hline
\multicolumn{4}{c}{PL fit}\\\hline
$\Gamma$      &   $2.20\pm0.06$   &  $2.18\pm0.09$      &   $2.17\pm0.07$  \\
$f_{\rm ph}$\tablenotemark{a}  & $(1.45\pm0.14)\times10^{-8}$   &   $(9.77\pm1.20)\times10^{-9}$  &  $(1.02\pm0.13)\times10^{-8}$   \\
TS            &  330    &    303    &  192   \\\hline
\multicolumn{4}{c}{PLE fit}\\\hline
$\Gamma$   &   $0.68\pm0.32$    &   $0.96$ (fixed)    &  $1.26\pm0.27$   \\
$E_{\rm cutoff}\tablenotemark{b}$ &  $1.16\pm0.27$    &  $1.21\pm0.12$    &  $2.05\pm0.71$   \\
$f_{\rm ph}$\tablenotemark{a} &  $(1.01\pm0.14)\times10^{-8}$   &  $(8.29\pm0.79)\times10^{-9}$    &  $(8.42\pm1.24)\times10^{-9}$  \\
TS            &  403   &    368    & 219  \\
\enddata
\tablenotetext{a}{Photon flux in unit of photons~cm$^{-2}$~s$^{-1}$ measured in the range of 0.2-300~GeV.}
\tablenotetext{a}{Cut-off energy in unit of GeV.}
\label{spec_par}
\end{deluxetable}
\end{center}

\acknowledgments{
The authors would like to thank Paul Ray and the anonymous referee for providing a code for computing phaseogram and useful 
comments for improving the quality of this manuscript. 
This project is supported by the National Science Council of the
Republic of China (Taiwan) through grant NSC100-2628-M-007-002-MY3 and
NSC100-2923-M-007-001-MY3. 
CYH is supported by the National Research Foundation of Korea through grant 2011-0023383.
A.K.H.K. gratefully acknowledges support
from a Kenda Foundation Golden Jade Fellowship. J.T. and K.S.C. are supported by a GRF grant of HK Government under HKU700911P.
}

\end{document}